\documentstyle[12pt]{article}
\begin{document}
\date{}
\author{S. M. Troshin, N. E. Tyurin\\
 Institute for High Energy Physics,\\ 142284 Protvino, Moscow
Region, Russia}
\title{Theoretical Aspects of Single-Spin Asymmetries Studies}
\maketitle

\begin{abstract}
We consider theoretical background for experimental measurements of
single-spin asymmetries. We stress the non-perturbative QCD aspects of
observed asymmetries in hadronic reactions.
\end{abstract}

The very important direction in spin studies is connected with the
long--standing problem of one--spin transverse asymmetries
observed in violent hadron reactions \cite{krsh}, \cite{helr}. It is well
known fact that the experimental data manifest significant
one--spin transverse asymmetries.

For example, the behavior of analyzing power in hadronic scattering
is rather surprising. Indeed, we could expect significant spin
effects in soft reactions where the chiral $SU(3)_L\times SU(3)_R$
symmetry of QCD Lagrangian is spontaneously broken down to $SU(3)_V$
and therefore, there is no ground for helicity conservation. However, the
observed analyzing power in the region of low transferred momenta is small
and decreases with energy like an inverse power of energy.

On the other side, contrary to our QCD expectations analyzing power
increases with transverse momentum when we trying to explore the
region of short distances. In this kinematical region we
should observe helicity conservation due to chiral invariance of
 QCD Lagrangian. Hadron helicity conservation in hard
processes is a general principle of perturbative QCD.
Violation of this principle have been observed in elastic
 $pp$--scattering, in two-body hadronic decays of
$J/\psi$  and there are also indications for such violation
in the measurements of Pauli form factor $F_2(Q^2)$.

It is evident now that new ideas and experimental
data are urgently needed to study  dynamics of the spin effects.

We consider possible dynamical mechanism of spin effects in elastic
scattereing.
In Ref. \cite{nuov} we
used the notions of effective chiral quark model
 for the description of elastic
scattering at small and large angles.
Different aspects of hadron dynamics were accounted in
the framework of effective
Lagrangian presented as a sum of three terms:
\begin{equation}
{\cal{L}}={\cal{L}}_\chi +{\cal{L}}_I+{\cal{L}}_C.
\end{equation}
 ${\cal{L}_\chi }$ is the term responsible for the spontaneous chiral
symmetry breaking:
\begin{equation}
{\cal{L}}_\chi =\bar{\psi }(i\partial ^\mu \gamma _\mu -\hat{m})\psi
+{\cal{L}}_4+{\cal{L}}_6.
\end{equation}
${\cal L}_4$ is the NJL four-fermion interaction, ${\cal L}_6$  is
the $U_A(1)$--breaking 6--quark interaction.
${\cal L}_\chi $ is responsible for providing  constituent  quark
masses and for the structure of constituent
quark which includes valence  quark
and cloud of quark-antiquark pairs \cite{stwi}.
${\cal  L}_I$  describes the  interaction   of
constituent  quarks and ${\cal  L}_C$ --- their  confinement.
These parts of effective interaction were  taken  into  account  at
phenomenological level.

In such a model quarks  appear as quasiparticles  and  have  a  complex
structure.
Besides its mass (consider $u$-quark as an example)
\begin{equation}
m_u = m^0_u-g_4 \langle uu \rangle - g_6 \langle\bar d
d\rangle\langle\bar s s\rangle
\end{equation}
the constituent quark has a finite size. We assume that the  strong
interaction radius of $q$-quark $r_q$ is determined by its mass:
$r_q = \xi /m_q$.
The common feature of the chiral models  is the representation
 of a baryon as an inner core carring the baryonic charge  and
an outer condensate surrounding this core \cite{isl}. Following
this picture it
is natural to represent a hadron  consisting  of  the  inner  region
where valence quarks are located and the outer region filled  with
quark condensate \cite{nuov}.
Such a picture for the hadron structure implies  that  overlapping  and
interaction of peripheral condensates at hadron  collision  occurs  at
the first stage. In the overlapping region the condensates
interact and as a result the massive quarks appear.
Being released the  part  of  hadron  energy  carried  by  the
peripheral condensates goes for the
generation of massive quarks. In another words  nonlinear field
couplings  transform  kinetic  energy  into  internal  energy  of
 dressed quarks (see the arguments for
this mechanism in \cite{carr} and references therein for the
earlier works). Of course, the number of such  quarks fluctuates.  The
average number of quarks should be proportional  to convolution  of
the condensate distributions $D^H_c$ of colliding hadrons:
\begin{equation}
N(s,b) \propto N(s)\cdot D^A_c \otimes D^B_c,
\end{equation}
where the function  $N(s)$  is  determined  by   the
thermodynamics  of
transformation of  kinetic  energy  of interacting  condenstates
to the internal energy of massive quarks. To estimate the
$N(s)$
it is feasible to assume that it is determined by the  maximal
possible energy dependence
\begin{equation}
N(s) \simeq \kappa  \frac{(1-\langle x\rangle _q)\sqrt{s}}{m_q},
\end{equation}
where $\langle x \rangle _q$ is the average fraction of energy carried by
valence  quarks, $m_q$ is the mass of constituent quark.

In the model \cite{nuov} valence quarks located in the central
part of a hadron are supposed to scatter in a
quasi-inde\-pen\-dent way  by  the
produced massive quarks at given impact parameter
 and by  the other valence
quarks. The averaged scattering  amplitude  of  valence
quark then may be represented in the form
\begin{equation}
\langle f_q (s,b) \rangle = [N(s,b) + N-1]\langle V_q(b) \rangle,
\end{equation}
where $N=N_1+N_2$  is  the  total  number  of  valence  quarks  in
colliding hadrons, and $\langle V_q(b) \rangle$ is  the  averaged
amplitude  of single quark-quark scattering \cite{nuov}.

In this approach elastic scattering  amplitude satisfies  unitarity
equation since it is constructed as a solution  of  the  following
equation \cite{logu}
\begin{equation}
F = U + iUDF \label{xx}
\end{equation}
which is presented here in operator form. This relation  allows  one
 to satisfy unitarity provided the inequality $\mbox{Im} U(s,b) \geq 0$
is fulfilled. The function $U(s,b)$  (generalized  reaction
matrix) \cite{logu} --- the basic dynamical quantity of  this
approach
--- is chosen as a product of the averaged quark amplitudes
\begin{equation}
U(s,b) = \prod^{N}_{q=1} \langle f_q(s,b)\rangle
\end{equation}
in accordance  with assumed quasi-independent  nature  of  valence  quark
scattering.

The $b$--dependence of function $\langle f_q \rangle$
is related to the quark formfactor behavior
$\propto ({\vec{q}}^2+m_q^2/\xi ^2)^{-2}$ and has a simple form
\cite{nuov}
$\langle f_q \rangle\propto\exp(-m_qb/\xi )$.

Following the lines of the above considerations, the generalized reaction
matrix in the pure imaginary case can be represented in the form
\begin{equation}
U(s,b) = iG(N-1)^N \left [1+\alpha \frac{\sqrt{s}}{m_q}\right]^N
\exp(-Mb/\xi ), \label{x}
\end{equation}
where $M =\sum^N_{q=1}m_q$. This expression allows one to
get the scattering amplitude  as  a  solution  of  Eq. \ref{xx}  which
reproduces the main regularities observed in  elastic  scattering
at small and large angles and consider spin phenomena.

For that purposes
system of equations for helicity amplitudes has been solved and dynamical
mechanism of quark scattering with and without helicity flip has been
 considered.

In particular spin of constituent quark in this model comes from the orbital
moment of cloud of quark-antiquark pairs while the polarization of
valence current quark and the polarization of the cloud
 of $\bar q q$ pairs compensate each other,
e.g. for $z$--component of spin it means
\[
S_q=S_{q_V}+S_{\{\bar q q\}}+\langle L_{\{\bar q q\}}\rangle
=1/2-1/2+1/2=1/2.
\]
The above compensation occurs due to account of axial $U(1)_A$
anomaly in the framework of effective QCD.
While considering the constituent quark as an extended object we
 can represent its spin as follows:
\[
S_q= \langle L_{\{\bar q q\}} \rangle =\omega I_q,
\]
where $\omega$ is the angular velosity of quark matter inside the
 constituent quark and $I_q$ its moment of inertia. These notions
on spin of constituent quark follows from consideration of spin in the
framework of effective lagrangian approach to QCD \cite{frit}, \cite{ellis}.

It should be noted that since spin of constituent quark is due to
its orbital angular momentum the corresponding wave function should
be equal to zero at $r=0$ due to centrifugal barrier.
Such picture was advocated for the proton as a whole
 by Ralston and Pire \cite{rlst}.

Quark helicity flip in the model is provided by the mechanism of quark exchange
where valence quark is exchanged with the quark produced under interaction
of condensates. These quarks have different helicities and therefore such
mechanism can lead to helicity flip quark scattering. Helicity non-flip
quark scattering has another origin resulting from optical type of interaction.
The above difference of these mechanisms leads to the different energy
dependence and different phases of helicity flip and non-flip quark scattering
amplitudes.
Helicity amplitudes at hadron level in this approach as it was already
mentioned are obtained as solutions
of coupled system of equations which accounts unitarity in direct
channel.
 Analyzing power in
the framework of this model does not decrease with energy and
has a non--zero value at $s\rightarrow \infty  $. The value of
analyzing power depends on the fraction of energy carried by
valence quarks $k$ and the phase difference $\Delta (s)\propto
(1-k)\sqrt{s}/m_q $ and has the following form
\begin{equation}
A(s,\theta )=\frac{4\sin
\Delta (s)}{(1-k)N}f(\theta )\left[1+O\left(\frac{m_q^2}{s}\right)\right],
\end{equation}
where $N$ is the total number of valence quarks in colliding
hadrons and $f(\theta)$ is the known function of scattering
angle. Asymmetry here results from interference of helicity amplitudes
which occurs due to resonance type of quark helicity flip scattering
and continuum type of quark helicity non-flip scattering.

Analyzing power at $\sqrt{s}= 2$ TeV and $-t=10$ $(GeV/c)^2$ in
$p_{\uparrow}\bar{p}$--elastic scattering is predicted to be $12
\%$ while at $-t=5$ $(GeV/c)^2$ --- $A=7\%$. Other non--perturbative models
\cite{mqm}, \cite{diq} also predict
non--zero values for analyzing power in TeV energy range.

To summarize, the measurement of analyzing power in elastic
scattering at high energies will allow
\begin{itemize}
\item
test perturbative QCD,
mechanism, get knowledge on the region of applicability of
perturbative QCD, study the transition from
nonperturbative to perturbative phase of QCD;
\item
study of hadron structure and non--perturbative effects:
spontaneous breaking of chiral symmetry and confinement.
\end{itemize}

\section*{Acknowledgements}
We are grateful to N. Akchurin, M. Anselmino, A. D. Krisch and J. P. Ralston
for interesting discussions, one of us (S.T.) express also his gratitude
to A. D. Krisch for support of the visit to this Simposium.

\end{document}